\documentclass[sigconf,nonacm]{acmart}
\usepackage{algorithm}
\usepackage{algpseudocode}
\usepackage{booktabs}
\usepackage{multirow}
\usepackage{array} 
\usepackage{caption}
\usepackage{subcaption}

\usepackage{xcolor}

\def\model/{\textsc{LightLM}}

\AtBeginDocument{%
  \providecommand\BibTeX{{%
    \normalfont B\kern-0.5em{\scshape i\kern-0.25em b}\kern-0.8em\TeX}}}

\setcopyright{acmcopyright}
\setcopyright{rightsretained}
\copyrightyear{2023}
\acmYear{2023}

\begin{document}




\title{LightLM: A Lightweight Deep and Narrow Language Model for Generative Recommendation}


\author{Kai Mei}
\affiliation{%
  \institution{Rutgers University}
  \institution{New Brunswick, NJ, US}
  \country{}
}
\email{kai.mei@rutgers.edu}

\author{Yongfeng Zhang}
\affiliation{%
  \institution{Rutgers University}
  \institution{New Brunswick, NJ, US}
  \country{}
}
\email{yongfeng.zhang@rutgers.edu}

\renewcommand{\shortauthors}{Kai Mei and Yongfeng Zhang}

\begin{abstract}
This paper presents \model/, a lightweight Transformer-based language model for generative recommendation. While Transformer-based generative modeling has gained importance in various AI sub-fields such as NLP and vision, generative recommendation is still in its infancy due to its unique demand on personalized generative modeling. Existing works on generative recommendation
often use NLP-oriented Transformer architectures such as T5, GPT, LLaMA and M6, which are heavy-weight and are not specifically designed for recommendation tasks. 
\model/ tackles the issue by introducing a light-weight deep and narrow Transformer architecture, which is specifically tailored for direct generation of recommendation items. 
This structure is especially apt for straightforward generative recommendation and stems from the observation that language model does not have to be too wide for this task, as the input predominantly consists of short tokens that are well-suited for the model's capacity. We also show that our devised user and item ID indexing methods, i.e., Spectral Collaborative Indexing (SCI) and Graph Collaborative Indexing (GCI), enables the deep and narrow Transformer architecture to outperform large-scale language models for recommendation. Besides, to address the hallucination problem of generating items as output, we propose the constrained generation process for generative recommenders.
Experiments on real-world datasets show that \model/ outperforms various competitive baselines in terms of both recommendation accuracy and efficiency.
The code can be found at \href{https://github.com/dongyuanjushi/LightLM}{https://github.com/dongyuanjushi/LightLM}.

\end{abstract}




\maketitle

\section{Introduction}

Generative recommendation~\cite{liang2018variational,p5,m6,wang2023diffusion,chen2022generative,liu2021variation,yuan2019simple} gains momentum in recent years. Previous discriminative recommenders calculate user-item scores one by one and then create the ranking list. In contrast, generative recommenders aim to directly generate the items for the given user, avoiding the inefficient one-by-one score ranking process taken by discriminative recommenders.

The primary challenge in generative recommendation lies in the effective representation of users and items. To achieve this, the deployment of unique and efficient identifiers (IDs) is essential for both users and items, which helps to avoid the hallucitation problem \cite{alkaissi2023artificial,zhao2023survey,chang2023survey} when generating long-text item descriptions for recommendation. This is important for recommender systems since the generated item should be really existing items in the item database \cite{hua2023index,fan2023recommender,lin2023can,wu2023survey,liu2023pre}. It is important to mention that, ID in this context, is not confined to the embedding vectors employed in prior studies ~\cite{li2023exploring,fu2023exploring,yuan2023go}. Instead, it is a broader concept and its format can vary from being the item title, an embedding vector, or a sequence of tokens. In this study, our focus is primarily on utilizing sequences of tokens as IDs. The rationale behind this choice is that these token sequences are typically concise and can guarantee uniqueness. These qualities are especially vital for generative recommendations that operate without a second stage of ranking or retrieval. The unique, short token sequences enable efficient and effective generative recommendation.

\begin{figure*}[htbp]
    \centering
    \includegraphics[width=0.8\textwidth]{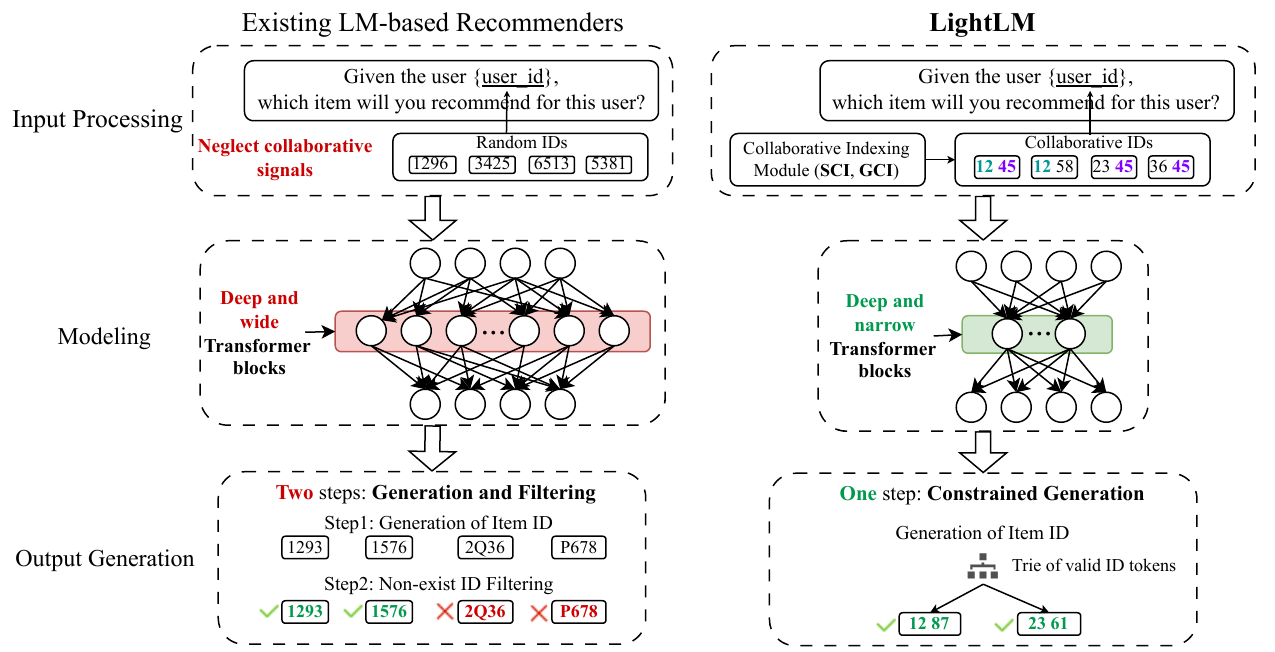}
    \vspace{-5pt}
    \caption{Overview of \model/, which presents the key differences between \model/ and existing LM-based recommenders in the aspects of input processing, modeling and output generation. Regarding the input processing, the same colored texts in different IDs represent the shared collaborative signals. Regarding the output generation, texts in green represent valid IDs which exist in data, while texts in red represent non-exist IDs.}\label{fig:overview}
    \vspace{-10pt}
\end{figure*}

Existing generative recommendation systems based on language models (LMs) are still in the nascent stage. Preceding studies on LM-based generative recommendation, such as P5~\cite{p5,xu2023openp5}, M6-Rec~\cite{m6}, InstructRec \cite{zhang2023recommendation} have treated recommendation as a task akin to natural language generation, employing NLP-focused Transformer architectures such as T5~\cite{raffel2020exploring}, GPT~\cite{brown2020language}, LLaMA \cite{touvron2023llama} and M6~\cite{lin2021m6}. However, these architectures, not being specifically optimized for recommendation tasks, may not fully harness the model's recommendation potential due to distinct characteristics that set recommendation tasks apart from natural language tasks. For instance, NLP Transformers generally process input sentences of variable and often substantial length, necessitating a considerable depth and width in their architecture. However, in the simple straightforward generative recommendation scenario, the model directly generates the recommended item IDs from the input user ID and such input only contains a few tokens. Furthermore, the primary goals of NLP and recommendation tasks diverge. While natural language tasks prioritize the fluency and diversity of generated results, recommendation tasks are more concerned with the precision of the recommended results.

More specifically, LMs for NLP tasks typically take the deep and wide design for Transformers.
Specifically, these LMs adopt larger inner dimensions for Feed-Forward (FF) layers than the dimension of attention layers. For example, if the dimension of the attention layer is $d$, then the inner dimension of Feed-Forward layer is usually $n$ times of the attention layer dimension, i.e., $d\times n$. Such design aims to enhance the representation capability as NLP corpora contains diverse tokens. By contrast, in the straightforward generative recommendation scenario, necessary tokens are much fewer than the NLP tasks. As a result, the Feed-Forward layers can be much narrower, such as $n$ fractions rather than times of the attention layer dimension (i.e., $d/n$), which enables efficient generation while acheiving better accuracy.

Inspired by this, we present \model/, a tailored LM-based recommender for generative recommendation, which narrows the inner dimension of LM while maintaining the original depth simutaneously. 
As illustrated in ~\autoref{fig:overview}, our recommender distinguishes from existing LM-based recommenders in three aspects.

(1) Regarding the input processing, we develop two advanced indexing methods, i.e., Spectral Collaborative Indexing (SCI) and Graph Collaborative Indexing (GCI), for capturing collaborative signals, which is more effective than the random indexing approach employed in previous recommenders. 

(2) Regarding the modeling, the input in generative recommendation demands only a few tokens, making it considerably shorter than typical NLP sentences. 
This means that traditional NLP Transformers may be over-parameterized for recommendation tasks. 
To address this issue, we propose to leverage deep and narrow Transformer blocks to replace the original deep and wide Transformer blocks, which reduce the inner dimension of all the Feed-forward layers. 
We choose to optimize the inner dimension of Feed-forward layers because the parameters of Feed-forward layers contribute the most to the overall parameters of Transformer blocks.

(3) Regarding the output generation, In the first phase, these systems produce item IDs indiscriminately, often resulting in the creation of non-existent or spurious IDs — a phenomenon termed the 'hallucination problem'. Given that such erroneous content can propagate misleading or even inappropriate information, particularly in delicate scenarios, these traditional systems incorporate a secondary phase: filter out the non-existent IDs by querying the ID dictionary.
In contrast, our approach focuses on accurately generating item IDs in a single step by leveraging constrained generation. We maintain a Trie structure to store valid ID tokens after tokenization, and employ this Trie as a constraint to prune the beam search tree. By doing so, we merge the two-step generation into one step without further filtering, thereby enhancing efficiency and avoiding hallucination issues.


The main contributions of this work to the community can be summarized as the following:

\begin{itemize}
    \item We propose \model/, a tailored Transformer-based recommender, which is effective and efficient for straightforward generative recommendation.
    \item Two advanced ID indexing methods, i.e., Spectural Collaborative Indexing (SCI) and Graph Collaborative Indexing (GCI), are devised to capture collaborative signals, thus enpowering \model/ for effective generation.
    \item We address the output hallucination problem by proposing a constrained generation technique for \model/.
    \item Experiments on various real-world datasets demonstrate that \model/ outperforms competitive baselines in terms of both recommendation accuracy and efficiency.
\end{itemize}


\section{Related Work}

\subsection{Discriminative Recommender}
Discriminative recommmenders~\cite{koren2009matrix,rendle2012bpr,he2017neural,wang2019neural,he2020lightgcn,mao2021simplex} mainly adopt collaborative filtering (CF) methods to model user-item interactions. During inference, the user-item scores are calculated one-by-one for each candidate item, and the scores are used to rank the candidate items for creating the recommendation list.
Such recommenders convert user and item representation into latent features (a.k.a. embeddings) and then apply different approaches on these embedding vectors to model user-item interactions. 
Matrix Factorization (MF)~\cite{koren2009matrix} determines the similarity between users and items through the dot product of their embeddings. BPRMF~\cite{rendle2012bpr} further enhances MF by introducing the Bayesian Personalized Ranking (BPR) loss to consider user's preference on interacted items over non-interacted items.
Later discriminative recommenders propose more complex structures for interaction modeling.
LightGCN~\cite{he2020lightgcn} refines the architecture of GCN by removing feature transformations and non-linear activation function to improve efficiency. Recformer~\cite{li2023text} adopts text embeding as user and item represntations for user-item matching. 
BERT4Rec~\cite{sun2019bert4rec} leverages Transformer-based LM to learn item representations and user history sequence representations for recommendation. S3Rec~\cite{zhou2020s3} devise auxiliary self-supervised objectives to learn the correlations among users, items and sequences to improve sequential recommendation.
Overall, they all follow the user-item matching score calculation paradigm, though using different methods for representing users and items.


\subsection{LLM-based Generative Recommender}

\noindent
Large Language Model (LLM) based generative recommenders mainly preprocess user-item interactions as sequences and then fine-tune a pre-trained LM for directly generating the recommended items ~\cite{p5,m6,zhang2023recommendation}.
For example, P5~\cite{p5} proposes a framework to convert multiple recommendation tasks into a unified ``prompt then predict'' pipeline and adopts multi-task optimization to train the personalized recommendation LM for generative recommendation. M6-Rec~\cite{m6} converts various recommendation tasks into natural language generation tasks and develops a parameter caching mechanism to avoid repeated computation during online inference. InstructRec \cite{zhang2023recommendation} unifies both recommendtion and search tasks into prompts and fine-tune a language model for generating the search or recommendation results. 
More relevant research on LLM-based recommender system can be seen in several recent surveys on the topic \cite{li2023large,training,wu2023survey,fan2023recommender,Industry,chen2023large}.






\section{Problem Definition} \label{sec:problem}
In this section, we define the straightforward recommendation task. We start by considering a user set \(\mathcal{U} = \{u_{1}, u_{2}, ..., u_{m}\}\) comprising \(m\) users and an item set \(\mathcal{I} = \{i_{1}, i_{2}, ..., i_{n}\}\) comprising \(n\) items. 
For each user \(u_{j}\) where \(j \in [1, m]\), there exists an interaction history characterized by a subset \(\mathcal{I}^{u_{j}} \subseteq \mathcal{I}\). The primary objective of a recommender is to recommend items not yet interacted by the user, defined as:
\begin{equation} \label{eq:1}
    \mathcal{R}^{u_{j}} = \mathcal{F}(\mathcal{I} \setminus \mathcal{I}^{u_{j}}|\mathcal{I}^{u_{j}}, \mathcal{\theta})
\end{equation}
As depicted in \autoref{eq:1}, a recommender utilizes the interaction history \(\mathcal{I}^{u_{j}}\) of user \(u_j\) to recommend items from the set \(\mathcal{I} \setminus \mathcal{I}^{u_{j}}\), which omits items already interacted from the complete item set. Here, $\mathcal{R}^{u_{j}}$ denotes the recommended items for user $u_{j}$, \(\mathcal{F}\) symbolizes the recommendation model, and \(\mathcal{\theta}\) represents its parameters.
It is crucial to understand that straightforward recommendation assumes the user history solely provides direct interaction data between users and items, without factoring in chronological interaction details or any additional metadata about users or items. Consequently, this form of recommendation becomes a purely ID-based scenario, designed to investigate the capabilities of recommenders when operating with minimal available information.

\section{Design of \model/}
In this section, we present the design of \model/. We begin with introducing the architecture of our deep and narrow LM. And then we present the collaborative user and item indexing algorthms specified for enhancing the collaborative representation of \model/. Finally, we discuss our constrained generation method to address the hallucination issue of output generation.

\subsection{Deep and Narrow Architecture}
\begin{figure}[t]
    \centering
    \includegraphics[width=0.35\textwidth]{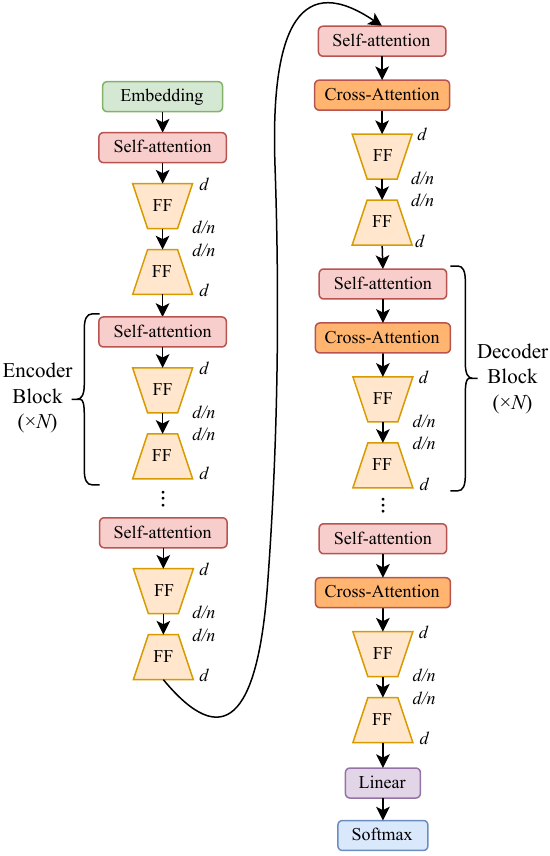}
    \caption{\model/, Deep and narrow encoder-decoder language model architecture, where we tailor the inner dimension $w$ from $d\times n$ to $\frac{d}{n}$ for all the feed-forward layers.}\label{fig:model_arc}
    \vspace{-10pt}
\end{figure}
We focus our exploration on refining the architecture of the encoder-decoder language model, as depicted in~\autoref{fig:model_arc}. To streamline our discussion and focus on the main aspects, we have omitted some other parts, such as positional encoding and normalization layers, which do not significantly impact the number of parameters in language models.
In language models such as BERT~\cite{devlin2018bert}, T5~\cite{brown2020language} and LLaMA \cite{touvron2023llama}, the Feed-Forward (FF) layers typically have a larger inner dimension compared with the default dimension used for attention operation. 
As is shown in~\autoref{fig:model_arc}, if the dimension of the input and output of the Feed-Forward layer is $d$, then the inner dimension of the Feed-Forward layer will be $d\times n$, where $n=4$ in standard Transformer \cite{vaswani2017attention}. This design choice aims to enhance the representation capabilities of the Feed-Forward layers, which is crucial for various natural language processing tasks.

In contrast, in recommendation tasks, the input typically consists of only a few natural language tokens, far fewer than the extensive NLP corpora. As a result, the conventional approach of increasing the inner dimension in the Feed-Forward layers might not be benefit recommendation tasks and such high-dimensional Feed-Forward layers can be the training bottleneck of model, since it contributes the most to the amount of model parameters. Therefore, we investigate narrowing down the Feed-Forward layers to meet the unique characteristics of recommendation tasks. Specifically, we tailor the inner dimension of the Feed-Forward layers $w$ from $d\times n$ to $\frac{d}{n}$ while maintaining the depth as other LMs do, which makes the \model/ model deep and narrow.

\subsection{User and Item Indexing} \label{sec:indexing}
In this section, we will present our indexing techniques to enhance \model/ for capturing collaborative signals in users and items.
\subsubsection{User and Item Graphs}
\begin{figure}[htbp]
    \centering
    \includegraphics[width=0.45\textwidth]{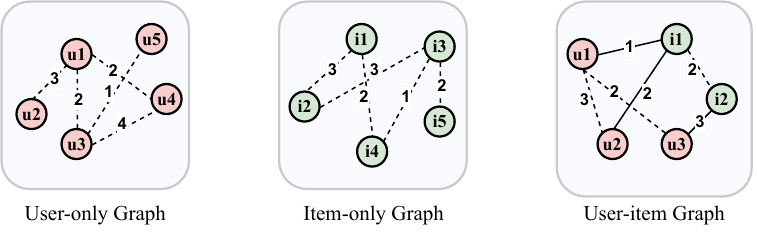}
    \caption{User and Item Graphs, where nodes in red represent users, nodes in green represent items, dotted lines represent co-occurrence times, solid lines represent interaction times.}\label{fig:graph}
    \vspace{-10pt}
\end{figure}

Before delving into the indexing techniques, we first explore three different graph settings: the user-only graph, the item-only graph, and the user-item graph, which are illustrated in ~\autoref{fig:graph}.
The \textbf{user-only graph} exclusively contains user nodes. Each edge between two users represents the frequency of their co-interaction, i.e., the number of items that both users interacted with.
Similarly, the \textbf{item-only graph} consists of item nodes, with edges connecting two items indicating their co-occurrence frequency, i.e., the freqency that two items co-appear in
the same user's interaction sequence. 
The \textbf{user-item graph} combines both users and items, representing their relationships through edges. Besides the user-to-user and item-to-item edges, additional edges are introduced between users and items. These new edges are labeled with the interaction times between each user and item. This way, the user-item graph captures not only the associations among users and items but also reflects the intensity of their interactions.

\subsubsection{Spectural Collaborative Indexing (SCI)} \label{sec:sci}
Following existing works~\cite{xu2023openp5, hua2023index}, we leverage spectral clustering to capture the collaborative signals between users and items. 
Specifically, we construct Laplacian matrix corresponding to the user-only, item-only and user-item graphs, respectively. 
Given the graph $\mathcal{G}$, we calculate the Laplacian matrix $\mathcal{L=D-A}$, where $\mathcal{D}$ is the diagonal matrix and $\mathcal{A}$ is the adjacency matrix of this graph.
Then we perform spectral clustering to ensure that users/items sharing more collaborative similarity will be grouped into the same cluster. 
We use the standard spectral clustering implementation in the Python scikit-learn package\footnote{\href{https://scikit-learn.org/stable/modules/generated/sklearn.cluster.SpectralClustering.html}{https://scikit-learn.org/stable/modules/generated/sklearn.cluster.SpectralClustering.html}}.
We refrain from delving into extensive details about the spectral clustering method, as it is a well-established clustering algorithm commonly covered in textbooks~\cite{leskovec2020mining}. 

\subsubsection{Graph Collaborative Indexing (GCI)} \label{sec:gci}
Inspired by quantization techniques~\cite{gray1984vector,rane2013quantized,guan2019post, gersho2012vector}, we transform graphs into embedding vectors. These vectors are subsequently quantized to create collaborative IDs.
We begin by discussing our chosen model for graph embedding training. For both user-only graph and item-only graph, we utilize a Graph Convolutional Network (GCN) \cite{kipf2017semi} with two graph convolutional layers to determine their representations. 
We optimize our model using the cross entropy loss described by:
\begin{equation} \label{eq:2}
    \mathcal{L} = -\sum_{c=1}^{N}y_{c}\log(\hat{y}_{c})
\end{equation}
where \(y\) signifies the ground-truth label of the node and \(\hat{y}\) represents the node's predicted label. The ground-truth label of a node is the cluster ID that the node belongs to, which is decided by the spectural clustering method in Section \ref{sec:sci}. More details of the clustering process will be introduced in the experiments.
For the user-item graph, to leverage the user-item interaction information, we follow BPRMF \cite{rendle2012bpr} to optimize each user-item pair's dot product. The associated training objective is:
\begin{equation} \label{eq:3}
    \mathcal{L} = - \log \sum_{u,i \in \mathcal{I}^u} \text{sigmoid}(u \cdot i_{pos}-u \cdot i_{neg}) \cdot \text{softmax}(u \cdot i_{neg})
\end{equation}

Here, \(i_{pos}\) signifies items interacted by the user, while \(i_{neg}\) refers to a sampled item that is not interacted by the user.
After obtaining the node embeddings, they are quantized into integer IDs \cite{guan2019post}. More specificially, we employ the K-Means clustering algorithm to cluster the embedding vectors into subgroups based on scikit-learn's K-Means implementation\footnote{\href{https://scikit-learn.org/stable/modules/generated/sklearn.cluster.KMeans.html}{https://scikit-learn.org/stable/modules/generated/sklearn.cluster.KMeans.html}}. It is crucial to note that we incorporate Z-score normalization over the embedding vectors~\cite{patro2015normalization}, as depicted in \autoref{eq:4}, prior to clustering. 
\begin{equation} \label{eq:4}
    Z = (X-\mu)/\sigma
\end{equation}
This process ensures that the embedding vectors are more distinctly separable \cite{patro2015normalization}. In this equation, \(X\) represents the original embedding vectors, while \(\mu\) and \(\sigma\) stand for the mean and standard deviation of \(X\), respectively.

\subsubsection{Hierarchical ID Construction}
We utilize a hierarchical approach to create a tree structure for indexing users/items in both SCI and GCI, as depicted in~\autoref{alg:SCI}. 
\begin{algorithm}
\small
\caption{Hierarchical ID Construction Algorithm} \label{alg:SCI}
\begin{algorithmic}
\Require Index dictionary $\mathcal{D}$, level one labels after the pre-clustering $\mathcal{L}_{1}$, the number of index levels $\mathcal{K}$, the number of clusters used in clustering $\mathcal{N}$, maximum number of entry (user/item) in each cluster $\mathcal{M}$.

\For{k $\gets$ 1\,to\,$\mathcal{K}$}

    \For{n $\in$ $[1, \mathcal{N}]$}
        \If{ number of $n$ in $\mathcal{L}_{k}$ > $\mathcal{M}$}
            \State $\mathcal{E}_{n}$ = [$e$ for $(e,s) \in \mathcal{L}_{k}$ if $s=n$]
            \State Build remap $\mathcal{P}$ for storing entry's order in the original list.
            \State $\mathcal{S}=clustering(\mathcal{E}_{n})$
            \For{$e,s \in \mathcal{E}_{n}, \mathcal{S}$}
                \State $\mathcal{D}[\mathcal{P}[e]] \leftarrow \mathcal{D}[\mathcal{P}[e]]+[s]$
                \State $\mathcal{L}_{k+1}.append(e,s)$
            \EndFor
        \EndIf
\EndFor
\EndFor
\State Deduplicate for $id \in \mathcal{D}$ which belong to the same final cluster.
\end{algorithmic}
\end{algorithm}
At its core, the method involves assessing the number of entries (users/items) in each cluster at the current level. If this count surpasses the threshold $\mathcal{M}$, we initiate an additional clustering phase, subdividing the entries from the current cluster into sub-clusters on the subsequent level. 
This recursive action concludes once every cluster at the present level has at most $M$ entries. 
Importantly, we employ an identical number of clusters, $N$, throughout all the clustering levels of each graph. 
Besides, to address situations where multiple entries might fall into a single final cluster, we incorporate an extra deduplication for the indexing dictionary.

\subsection{Constrained Generation}
Our constrained generation is essentially a pruned beam search process. In this section, we first delve into the conventional beam search and subsequently discuss our enhancements.
The conventional beam search can be described as follows:
Given an initial token set \( O = (o, p) \), where \( o \) symbolizes the token and \( p \) represents its probability, we denote the beam width by \( B \). 
Initially, we set the sequence set \( S = O \).
For each time step \( t = 1, 2, \ldots, T \), with \( T \) being the maximum sequence length the model can generate, the process calculates the probability distribution of potential succeeding tokens for each token \( (o, p) \in O \) using the generative function \( H \),
\begin{equation} \label{eq:5}
    P(o_{t}|s) = H(o_{t})
\end{equation}
where $o_{t} \in O, s \in S$
Subsequently, the sequence set can be expanded to \( S \) by supplementing each sequence \( (s, p) \in S \) with the \( B \) likeliest succeeding tokens as indicated by \( P(o_{t}|s) \):
\begin{equation} \label{eq:6}
    S = \left\{ (s \oplus o_t, p * P(o_t|s)) \mid o_t \in Top_B(P(o_t|s)) \right\}
\end{equation}
Here, \( \oplus \) represents concatenation, and \( Top_B(P(o_t|s)) \) indicates the top \( B \) tokens that will be concatenated with $s \in S$ based on \( P(o_t|s) \). After this, \( S \) is updated, and the expansion process continues.
Upon reaching the stopping criterion (such as reaching the maximum length \( T \)), the best \( B \) sequences in \( S \) are selected.

In the language model (LM) generation contexts, the initial set \( A \) comprises the entire vocabulary of the LM. Using this full vocabulary without any form of pruning introduces inefficiencies and potential inaccuracies. Specifically, it can be time-intensive, scanning the entire vocabulary to compute cumulative probabilities at every step. Additionally, it can introduce the hallucination issue in recommendations, where non-existent tokens are produced, diminishing recommendation precision.
To address this, we introduce our constrained generation methodology, which essentially prunes the traditional beam search. Recall that we previously obtained the collaborative ID tokens. Based on these ID tokens, we can build a hierarchical Trie\footnote{\href{https://en.wikipedia.org/wiki/Trie}{https://en.wikipedia.org/wiki/Trie}}. Each node layer in this Trie consists only of valid ID tokens at its respective construction level, as detailed in Section \ref{sec:indexing}.
Then the refined beam search can be represented as:
\begin{equation} \label{eq:7}
    S = \left\{ (s \oplus o_t, p * P(o_t|s,c)) \mid o_t \in Top_B(P(o_t|s,c)) \right\}
\end{equation}
Here, \( c \) signifies the Trie constraint, and \( a_t \) can only be derived from the leaf nodes of the current sequence \( s \)'s last token node. 
In this way, we can calculate cumulative probabilites at each step $t$ from the Trie at current node, which contains much fewer tokens than the fixed $O$.
This refinement notably diminishes the computational overhead associated with the conventional beam search and eliminates the generation's hallucination problem.

\section{Evaluation} \label{sec:eval}

\subsection{Experimental Settings} \label{sec:exp_setting}
Our experiments are conducted in Python 3.9 with PyTorch 1.13.1 and CUDA 11.4 on an Ubuntu 20.04 machine equipped with 8 NVIDIA RTX A5000 GPUs.

\noindent
\textbf{Datasets.}
Following existing works~\cite{kang2018self,zhou2020s3,xie2022contrastive,p5,xu2023openp5}, we conduct our experiments on Beauty and Toys, two commonly-used sub datasets from Amazon and the Yelp dataset. The Amazon datasets~\cite{ni2019justifying}
are sourced from Amazon.com\footnote{\href{https://cseweb.ucsd.edu/ jmcauley/datasets/amazon_v2/}{https://cseweb.ucsd.edu/jmcauley/datasets/amazon\_v2/}} for product recommendations, while the Yelp dataset\footnote{\href{https://www.yelp.com/dataset}{https://www.yelp.com/dataset}} provides a collection of user ratings and reviews for business recommendation. 
For a fair comparison, we utilize transaction records from January 1, 2019 to December 31, 2019 to preprocess, which is the same setting in previous works. 
Details of the dataset can be found in~\autoref{tab:dataset}.
We split the datasets into training, validation, and testing by
the frequently used leave-one-out setting: for each user’s interaction history, we put the second-to-last item into the validation
set, put the last item into the testing set, and construct training set using all the other items in the user's history. 
\begin{table}[htbp]
\centering
\footnotesize
\caption{Details of dataset, where rows 2-4 show the number of users, items and interactions, respectively, and row 5 shows the data sparsity.} \label{tab:dataset}
\begin{tabular}{@{}cccc@{}}
\toprule
Dataset        & Beauty  & Toys    & Yelp    \\ \midrule
\#Users        & 22361   & 19412   & 30431   \\
\#Items        & 12101   & 11924   & 20034   \\
\#Interactions & 198502  & 167597  & 316354  \\
Sparsity       & 99.93\% & 99.93\% & 99.95\% \\ \bottomrule
\end{tabular}
\end{table}

\noindent
\textbf{Baselines.}
To cover a wide scope of baselines as much as possible, we compare our method with both discriminative recommendation baselines (BPRMF~\cite{rendle2012bpr}, 
LightGCN~\cite{he2020lightgcn}, 
SimpleX~\cite{mao2021simplex}) and generative recommendation baseline (P5~\cite{p5}
). 

\noindent
\textbf{Evaluation Metrics.}
We use Hit Ratio at rank K (HR@K) and Normalized Discounted Cumulative Gain at rank K (NDCG@K) to evaluate recommendation performance. 
For a fair comparison, we obtain top-K recommended items from the whole item set for all the methods we evaluate, and we use K=5 and K=10 throughout the evaluation of this paper. 

\subsection{Implementation Details} \label{sec:implementation_details}
Regarding the indexing, we implement user-indexing (U) for user-only graph, item-indexing (I) for item-only graph, user-item indexing (UI) for both user-only and item-only graphs and user-item coindexing (CoUI) for useritem graph. Above notations are used throughout the experiments.
By default, we set the number of clusters \(N\) to 20 for SCI(U), SCI(I), SCI(UI), and \(N\) to 50 for SCI(CoUI). For GCI, across all four indexing settings, we use \(N = 20\) and set the embedding size \(E\) to 64. These choices are grounded in our experimental practice.

We construct the basic blocks of LightLM based on the transformers library\footnote{\href{https://github.com/huggingface/transformers}{https://github.com/huggingface/transformers}}. 
Specifically, we take the encoder-decoder architecture to build our model the same as previous works~\cite{p5} does.
 There are 6 layers for both encoder and decoder, the dimensions of embedding and self-attention layers are 512 and we use smaller dimension of Feed-Forward(FF) layers, which is different from the standard Transformer block. To faciliate training, we initialize the weights of layers except for FF layers from T5's pretrained weights.
For tokenization, we adopt the SentencePiece tokenizer~\cite{sennrich2015neural} with a vocabulary size of 32,128 to parse sub-word units. However, it is essential to note that we take special care with collaborative ID tokens, which are kept within the range of 1 to 999. To ensure that the collaborative ID tokens remain intact and are not further tokenized into subtokens, we add spaces between the tokens in an ID, for example, "13 25 46", instead of introducing extra tokens into the tokenizer vocabulary, which is different from~\cite{hua2023index,xu2023openp5}. The reason behind this approach is to prevent the SentencePiece tokenizer from breaking down the collaborative ID tokens into smaller units, as numbers from 1 to 999 already exist in the original vocabulary.
We randomly reinitialzie the embeddings of all the number tokens used in above indexing methods, which is inspired by~\cite{hua2023index}.

\begin{table*}[htbp]
\centering
\footnotesize
\setlength\tabcolsep{1pt}
\caption{Recommendation performance on straightforward recommendation tasks. Numbers in bold indicate the highest values, while underlined numbers denote the second highest values.} \label{tab:comparison_with_stoa}
\begin{tabular}{lcccccccccccccc}
\hline
\multirow{2}{*}{Method}          & \multicolumn{4}{c}{Beauty}                                            &  & \multicolumn{4}{c}{Toys}                                              &  & \multicolumn{4}{c}{Yelp}                                              \\ \cline{2-5} \cline{7-10} \cline{12-15} 
                                 & HR@5            & NDCG@5          & HR@10           & NDCG@10         &  & HR@5            & NDCG@5          & HR@10           & NDCG@10         &  & HR@5            & NDCG@5          & HR@10           & NDCG@10         \\ \hline
BPRMF                            & 0.0240          & 0.0150          & 0.0389          & 0.0198          &  & 0.0332          & 0.0179          & 0.0465          & 0.0242          &  & 0.0327          & 0.0219          & 0.0509          & 0.0277          \\
LightGCN                         & 0.0267          & 0.0165          & 0.0436          & 0.0219          &  & 0.0291          & 0.0187          & 0.0442          & 0.0233          &  & \underline{0.0619}    & \underline{0.0455} & 0.0827          & \underline{0.0522} \\
SimpleX                          & 0.0300          & 0.0180          & 0.0493          & 0.0243          &  & 0.0287          & 0.0163          & 0.0482          & 0.0238          &  & 0.0532          & 0.0353          & 0.0872          & 0.0465          \\
P5                               & 0.0317          & 0.0239          & 0.0437          & 0.0309          &  & 0.0261          & 0.0202          & 0.0335          & 0.0226          &  & 0.0404          & 0.0270          & 0.0615          & 0.0336          \\
\hline
\model/-SCI(U)    & \underline{0.0392}    & \underline{0.0305}    & 0.0522          & \underline{0.0347}    &  & \underline{0.0419}    & \underline{0.0323}    & \underline{0.0543}    & \underline{0.0364}    &  & 0.0440          & 0.0293          & 0.0651          & 0.0360          \\
\model/-SCI(I)    & 0.0142          & 0.0099          & 0.0224          & 0.0125          &  & 0.0207          & 0.0159          & 0.0317          & 0.0194          &  & 0.0136          & 0.0088          & 0.0230          & 0.0118          \\
\model/-SCI(UI)   & 0.0292          & 0.0207          & 0.0452          & 0.0259          &  & 0.0394          & 0.0334          & 0.0511          & 0.0377          &  & \textbf{0.0621} & 0.0402          & \underline{0.0972}    & 0.0480          \\
\model/-SCI(CoUI) & 0.0383          & 0.0279          & \textbf{0.0582} & 0.0342          &  & \textbf{0.0475} & \textbf{0.0330} & \textbf{0.0619} & \textbf{0.0376} &  & 0.0522          & 0.0321          & 0.0927          & 0.0450          \\ \hline
\model/-GCI(U)    & 0.0382          & 0.0287          & 0.0513          & 0.0321          &  & 0.0223          & 0.0148          & 0.0338          & 0.0203          &  & 0.0426          & 0.0285          & 0.0731          & 0.0386          \\
\model/-GCI(I)    & 0.0114          & 0.0114          & 0.0213          & 0.0148          &  & 0.0153          & 0.0118          & 0.0244          & 0.0168          &  & 0.0247          & 0.0188          & 0.0361          & 0.0249          \\
\model/-GCI(UI)   & 0.0348          & 0.0248          & 0.0445          & 0.0293          &  & 0.0298          & 0.0189          & 0.0441          & 0.0228          &  & 0.0561          & 0.0367          & \textbf{0.1012} & 0.0502    \\
\model/-GCI(CoUI) & \textbf{0.0431} & \textbf{0.0353} & \underline{0.0581}    & \textbf{0.0392} &  & 0.0412          & 0.0242          & 0.0528          & 0.0312          &  & 0.0618          & \textbf{0.0508}          & 0.0759          & \textbf{0.0553}          \\ \hline
\end{tabular}
\end{table*}

\subsection{Performance Comparison with State-of-the-Arts} \label{sec:compare_sota}
To maintain a fair comparison, we utilize the same dataset-split approach for all baselines and assess them using their standard parameters. We compare the baselines against eight variants of \model/ with different indexing configurations: For Spectral Collaborative Indexing, we have SCI(U), SCI(I), SCI(UI), and SCI(CoUI); while for Graph Collaborative Indexing, we consider GCI(U), GCI(I), GCI(UI), and GCI(CoUI). The comparative results are presented in \autoref{tab:comparison_with_stoa}. 
Across all three datasets, at least one variant of \model/ consistently surpasses the baselines, highlighting the effectiveness of our approach. In particular, 
\model/-GCI(CoUI) stands out on the Beauty dataset, outperforming both baselines and other \model/ variants.
For the Toys dataset, \model/-SCI(CoUI) achieves the highest recommendation accuracy across all metrics.
On Yelp, both \model/-SCI(UI) and \model/-GCI(UI) deliver superior results.
The advantage of indexing both user and item sides is evident, as it often provides richer collaborative context than single-sided indexing. 
Interestingly, on datasets like Toys, user-only indexing performs on par with more comprehensive useritem-indexing and useritem-coindexing. 
However, item-only indexing consistently lags behind in performance across all datasets. 
This can likely be attributed to many tasks primarily relying on user IDs, which necessitates a deeper collaborative context from the user side. 
Thus, item-only indexing struggles as it may not provide sufficient collaborative information for optimal generation.

\subsection{Ablation Studies}
In this section, we aim to investigate the influence of inner dimension $w$ of Feed-Forward layers and various indexing settings (SCI, GCI) on the recommendation performance. Throughout this section, we carry out experiments on the Toys dataset to assess the effects of these factors.

\subsubsection{Impact of different inner dimensions of Feed-Forward layers}
We investigated the influence of varying inner dimensions of Feed-Forward layers on recommendation performance. Specifically, we examined $w$ values of 16, 32, 64, and evaluated \model/ using the eight indexing configurations detailed in Section \ref{sec:compare_sota}. The results are shown in \autoref{tab:impact_of_inner}.
We do not observe a clear upward trend in recommendation accuracy with the increase in inner dimensions in \autoref{tab:impact_of_inner}. Notably, for configurations like GCI(UI) and GCI(CoUI), there's a decline in recommendation performance as the inner dimension rises. This implies that the Feed-Forward layers within Transformer blocks may not require excessively wide dimensions for generative recommendations, reinforcing the rationale behind our decision to tailor Transformers.
\begin{table*}[htbp] 
\centering
\footnotesize
\setlength\tabcolsep{1pt}
\caption{Impact of inner dimension $w$ of Feed-Forward layers on recommendation performance.} \label{tab:impact_of_inner}
\begin{tabular}{lcccccccccccccc}
\hline
\multirow{2}{*}{Model} & \multicolumn{4}{c}{Inner dimension $w=16$}             &  & \multicolumn{4}{c}{Inner dimension $w=32$}             &  & \multicolumn{4}{c}{Inner dimension $w=64$}             \\ \cline{2-5} \cline{7-10} \cline{12-15}
                         & HR@5   & NDCG@5 & HR@10  & NDCG@10 &  & HR@5   & NDCG@5 & HR@10  & NDCG@10 &  & HR@5   & NDCG@5 & HR@10  & NDCG@10 \\ \hline
\model/-SCI(U)           & 0.0419 & 0.0323 & 0.0543 & 0.0364  &  & 0.0521 & 0.0387 & 0.0610 & 0.0415  &  & 0.0489 & 0.0386 & 0.0600 & 0.0422  \\
\model/-SCI(I)           & 0.0208 & 0.0144 & 0.0340 & 0.0186  &  & 0.0217 & 0.0149 & 0.0307 & 0.0185  &  & 0.0139 & 0.0113 & 0.0230 & 0.0120  \\
\model/-SCI(UI)          & 0.0394 & 0.0311 & 0.0514 & 0.0377  &  & 0.0414 & 0.0314 & 0.0531 & 0.0352  &  & 0.0445 & 0.0330 & 0.0571 & 0.0371  \\
\model/-SCI(CoUI)        & 0.0475 & 0.0619 & 0.0330 & 0.0376  &  & 0.0520 & 0.0399 & 0.0655 & 0.0443  &  & 0.0504 & 0.0367 & 0.0656 & 0.0416  \\ \hline
\model/-GCI(U)           & 0.0382 & 0.0287 & 0.0513 & 0.0321  &  & 0.0381 & 0.0270 & 0.0512 & 0.0313  &  & 0.0395 & 0.0271 & 0.0516 & 0.0311  \\
\model/-GCI(I)           & 0.0114 & 0.0114 & 0.0213 & 0.0148  &  & 0.0153 & 0.0119 & 0.0234 & 0.0158  &  & 0.0139 & 0.0104 & 0.0202 & 0.0124  \\
\model/-GCI(UI)          & 0.0348 & 0.0248 & 0.0445 & 0.0293  &  & 0.0298 & 0.0189 & 0.0441 & 0.0228  &  & 0.0282 & 0.0202 & 0.0384 & 0.0235  \\
\model/-GCI(CoUI)        & 0.0431 & 0.0353 & 0.0581 & 0.0392  &  & 0.0412 & 0.0242 & 0.0528 & 0.0312  &  & 0.0332 & 0.0248 & 0.0503 & 0.0355  \\ \hline
\end{tabular}
\end{table*}
\subsubsection{Impact of different Spectral Collaborative Indexing settings} \label{sec:impact_sci}
\begin{figure*}[htbp]
    \caption{Impact of the number of clusters $N$ used in Spectral clustering on the recommendation performance under the SCI settings, where \autoref{fig:u_sci}, \autoref{fig:i_sci}, \autoref{fig:ui_sci}, \autoref{fig:coui_sci} refers to user-only indexing, item-only indexing, useritem indexing, useritem-coindexing, respectively.}
    \label{fig:sci}
    \begin{subfigure}[h]{0.24\textwidth}
    \includegraphics[width=\textwidth]{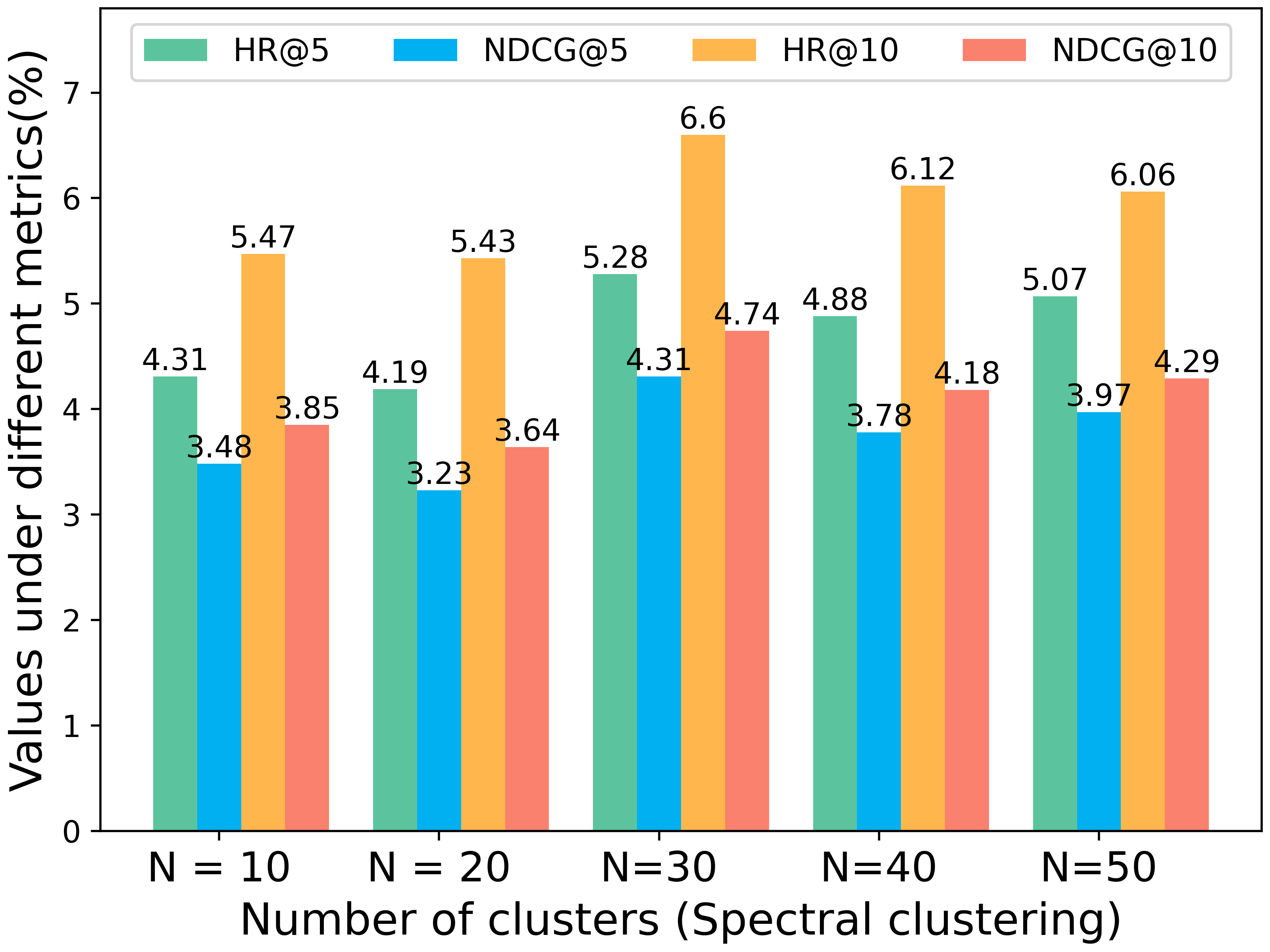}
        \caption{User-only indexing}
        \label{fig:u_sci}
    \end{subfigure}
    \hfill
    \begin{subfigure}[h]{0.24\textwidth}
    \includegraphics[width=\textwidth]{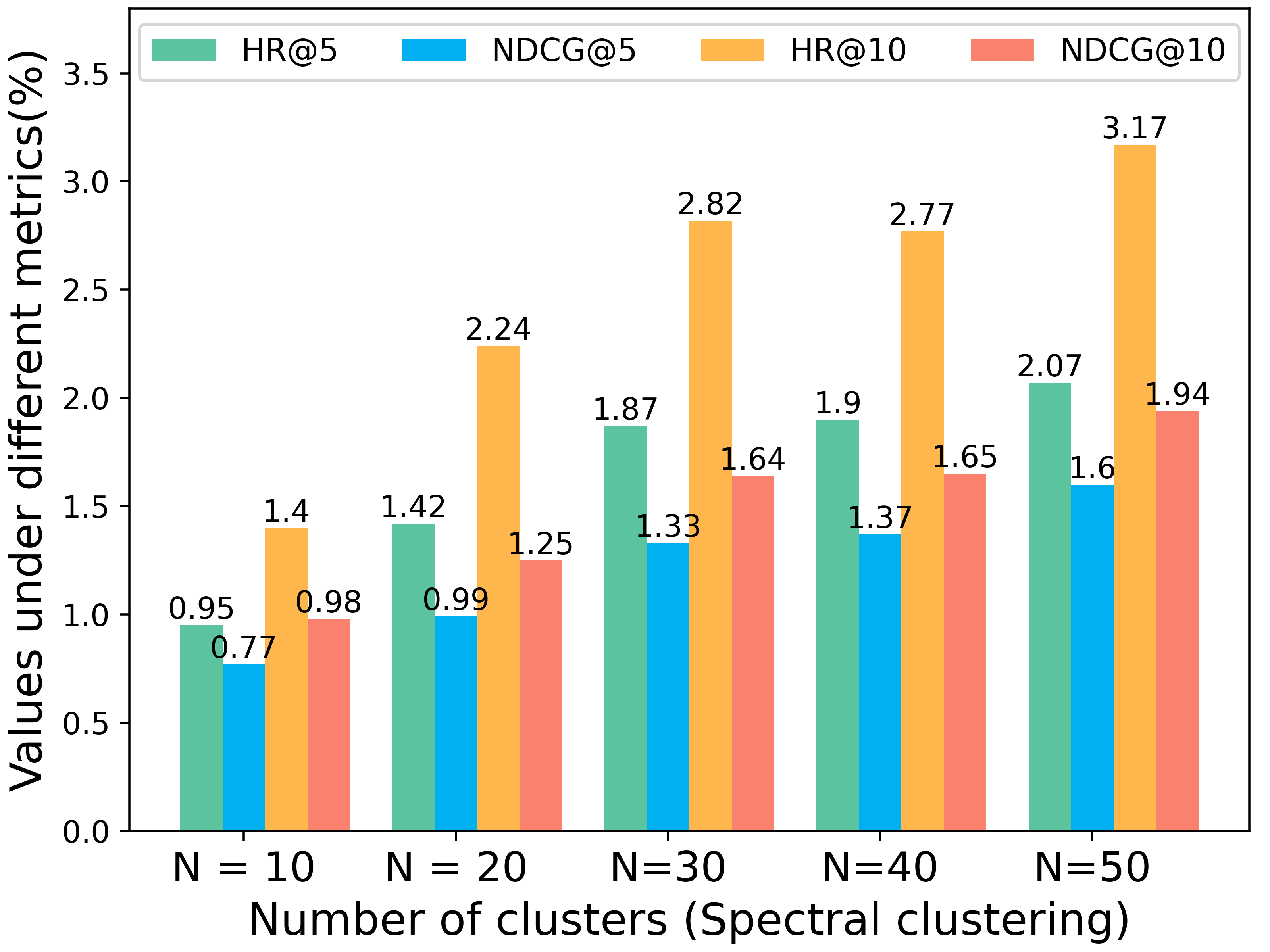}
        \caption{Item-only indexing}
        \label{fig:i_sci}
    \end{subfigure}
    \hfill
    \begin{subfigure}[h]{0.24\textwidth}
    \includegraphics[width=\textwidth]{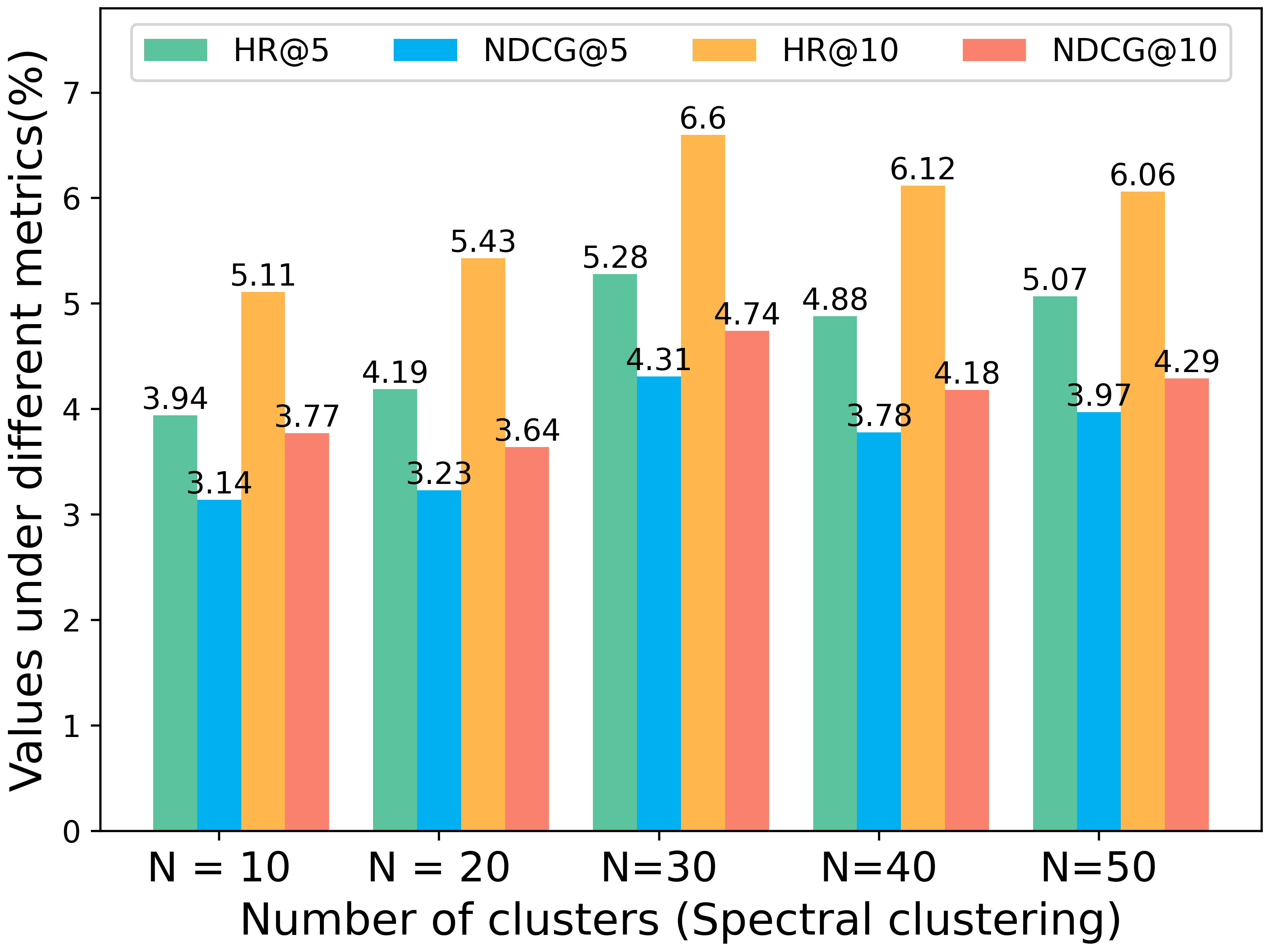}
        \caption{User-item indexing}
        \label{fig:ui_sci}
    \end{subfigure}
    \hfill
    \begin{subfigure}[h]{0.24\textwidth}
    \includegraphics[width=\textwidth]{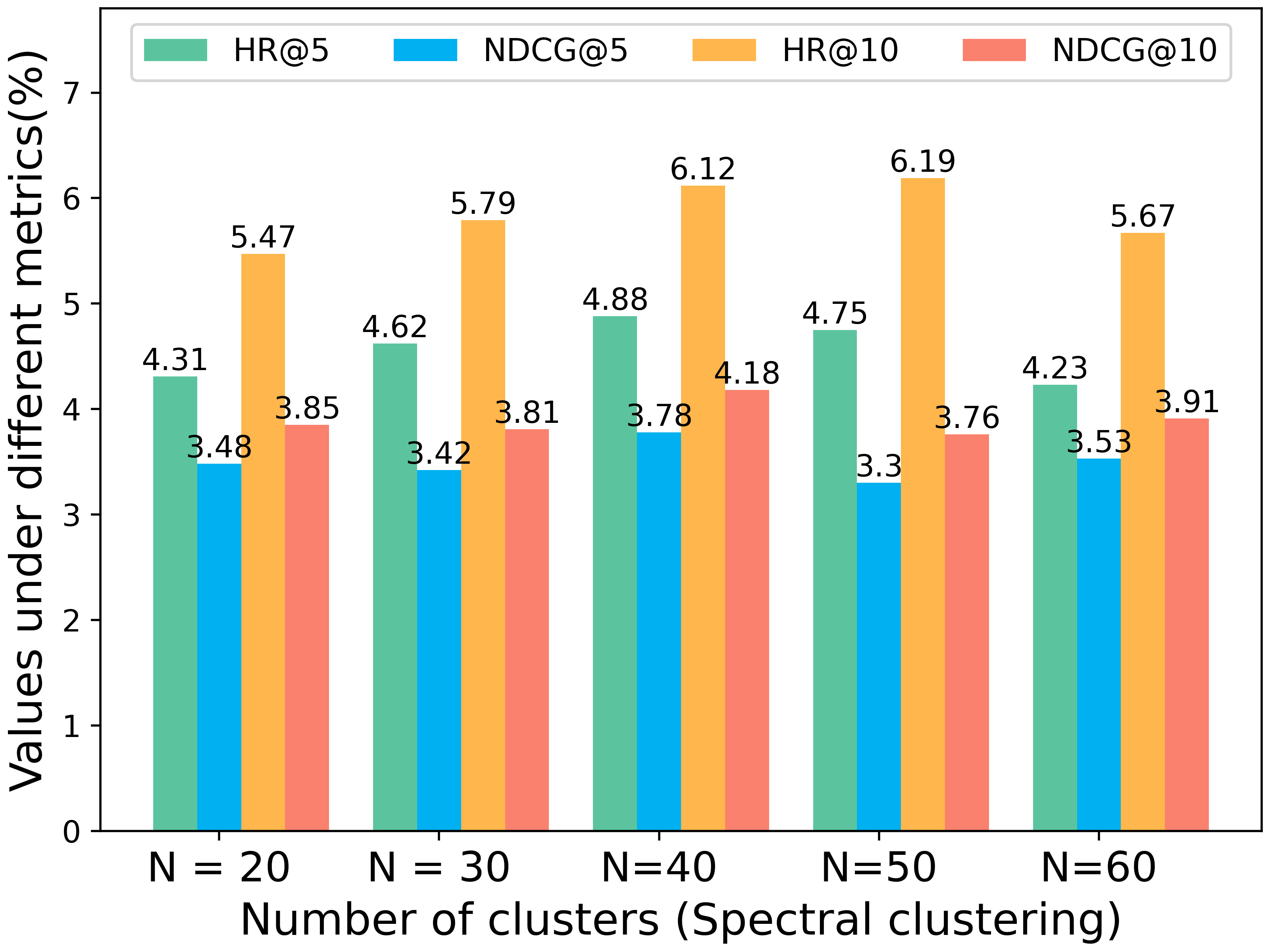}
        \caption{User-item coindexing}
        \label{fig:coui_sci}
    \end{subfigure}
    \hfill
\end{figure*}

We utilize the same number of clusters, denoted as $N$, across various levels for each indexing configuration, such as user-only or item-only. Our focus here is to examine how varying values of $N$ influence recommendation performance within the \textbf{SCI} framework, as depicted in \autoref{fig:sci}.
For SCI(U), SCI(I), and SCI(UI), we use $N$ values spanning from 10 to 50. However, for SCI(CoUI), we use $N$ ranging from 20 to 60. This differentiation is due to the fact that user-item coindexing necessitates a more extensive graph encompassing both user and item nodes.
From our observations, SCI(U) and SCI(UI) demonstrate optimal performance with $N$ set at 30 and 40. Meanwhile, for SCI(CoUI), peak performance emerges when $N$ is set to 40 and 50. This optimal performance can be attributed to the fact that at these $N$ values, the number of nodes that end up in the same final cluster is closer to $N$. This indicates a more balanced distribution of node numbers compared to other $N$ values may mitigate bias and augment recommendation precision. 

\subsubsection{Impact of different Graph Collaborative Indexing settings}
\begin{figure*}[htbp]
    \caption{Impact of the embedding dimension $E$ used in graph embedding on the recommendation performance under the GCI settings, where \autoref{fig:u_gci_e}, \autoref{fig:i_gci_e}, \autoref{fig:ui_gci_e}, \autoref{fig:coui_gci_e} refers to user-only indexing, item-only indexing, useritem indexing, useritem-coindexing, respectively.}
    \label{fig:gci_e}
    \begin{subfigure}[h]{0.24\textwidth}
    \includegraphics[width=\textwidth]{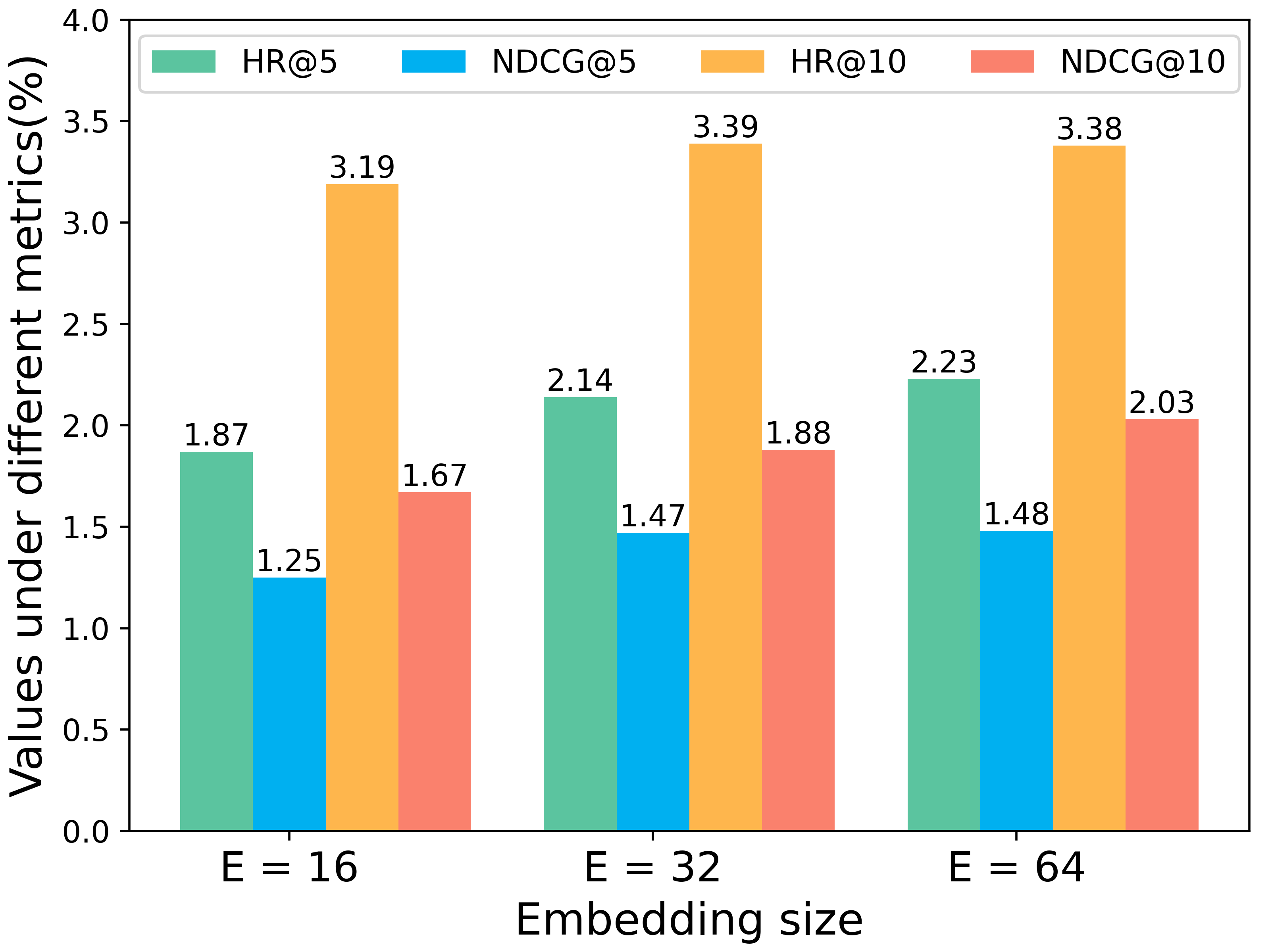}
        \caption{User-only indexing}
        \label{fig:u_gci_e}
    \end{subfigure}
    \hfill
    \begin{subfigure}[h]{0.24\textwidth}
    \includegraphics[width=\textwidth]{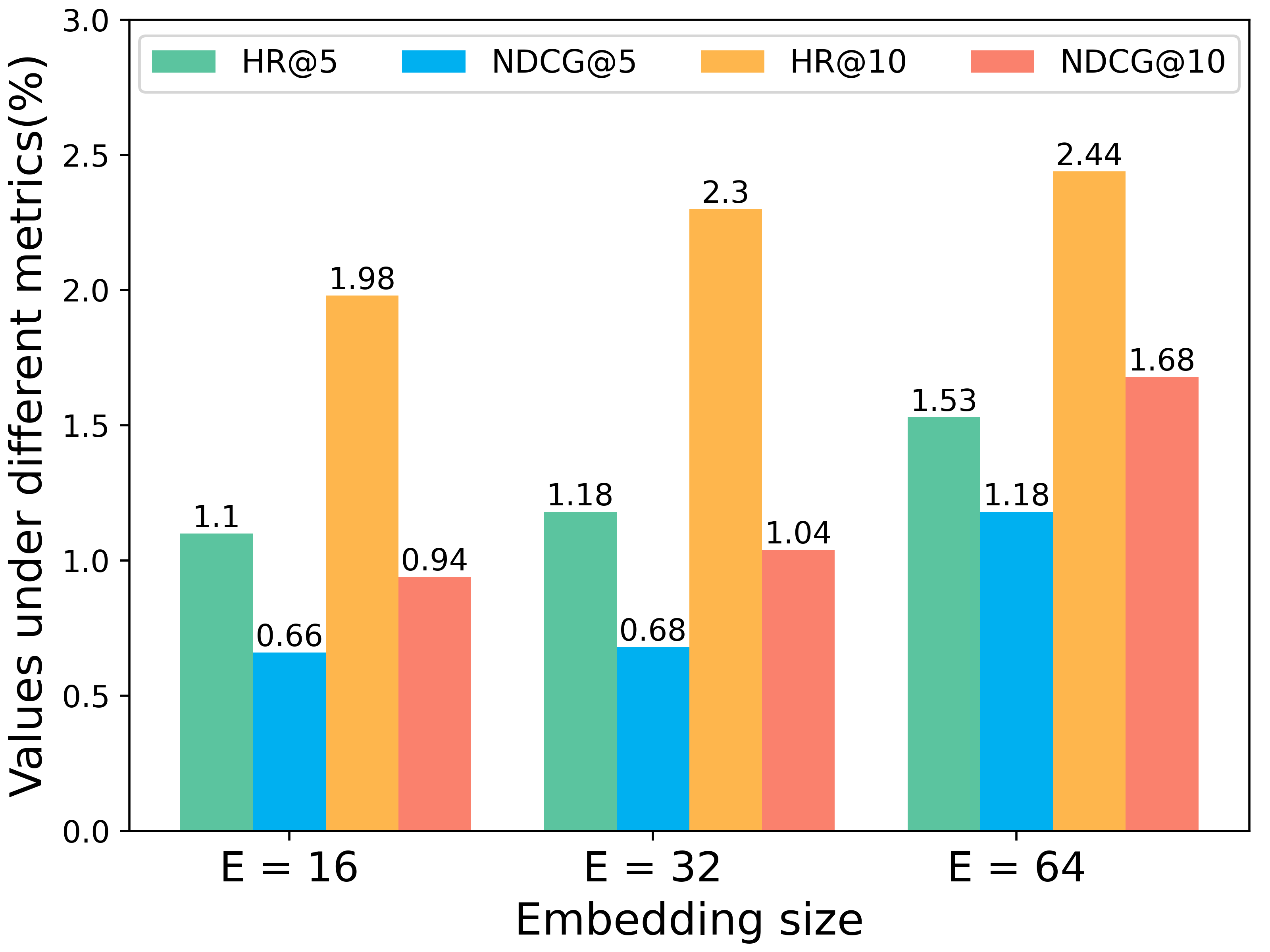}
        \caption{Item-only indexing}
        \label{fig:i_gci_e}
    \end{subfigure}
    \hfill
    \begin{subfigure}[h]{0.24\textwidth}
    \includegraphics[width=\textwidth]{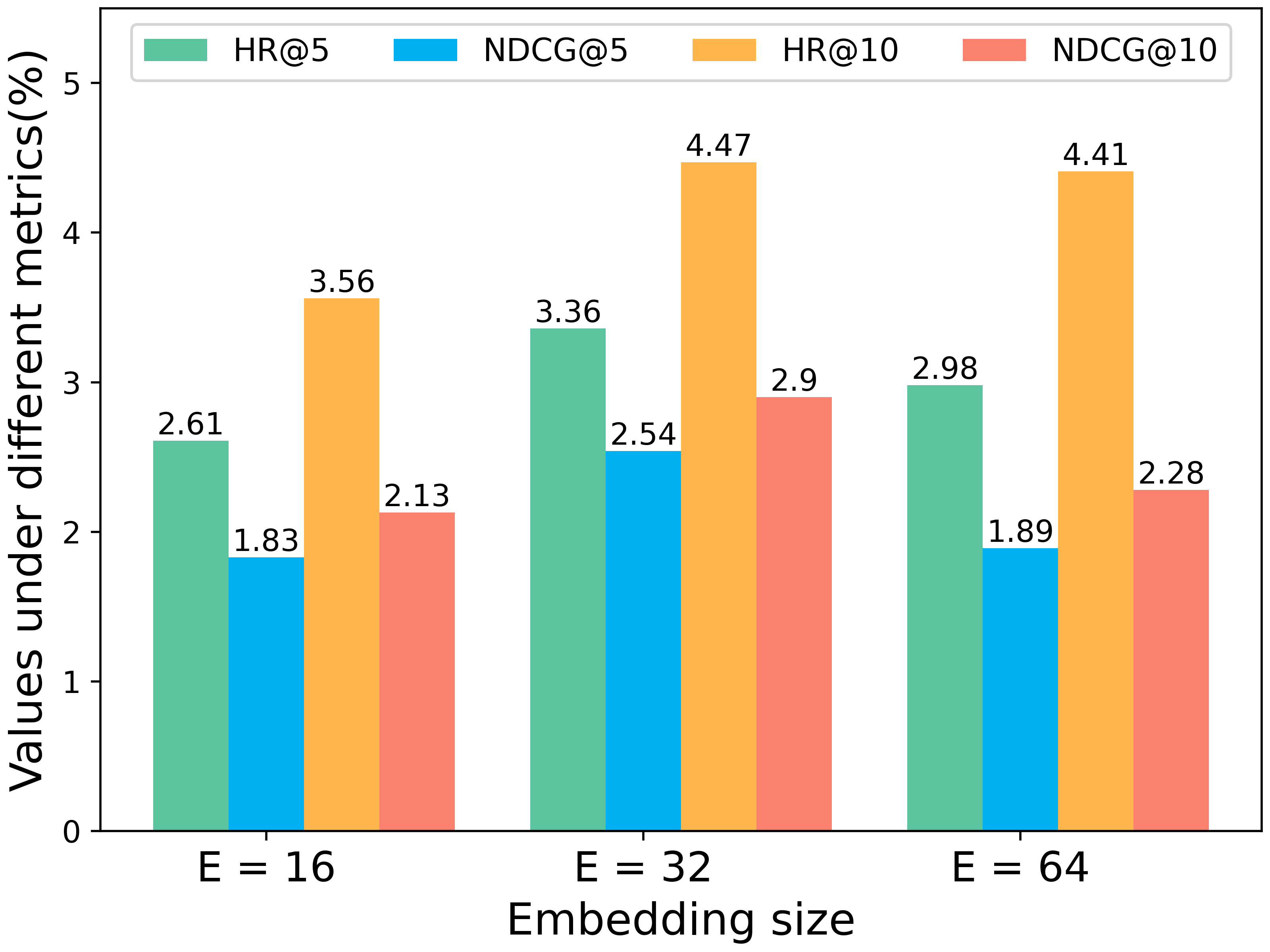}
        \caption{User-item indexing}
        \label{fig:ui_gci_e}
    \end{subfigure}
    \hfill
    \begin{subfigure}[h]{0.24\textwidth}
    \includegraphics[width=\textwidth]{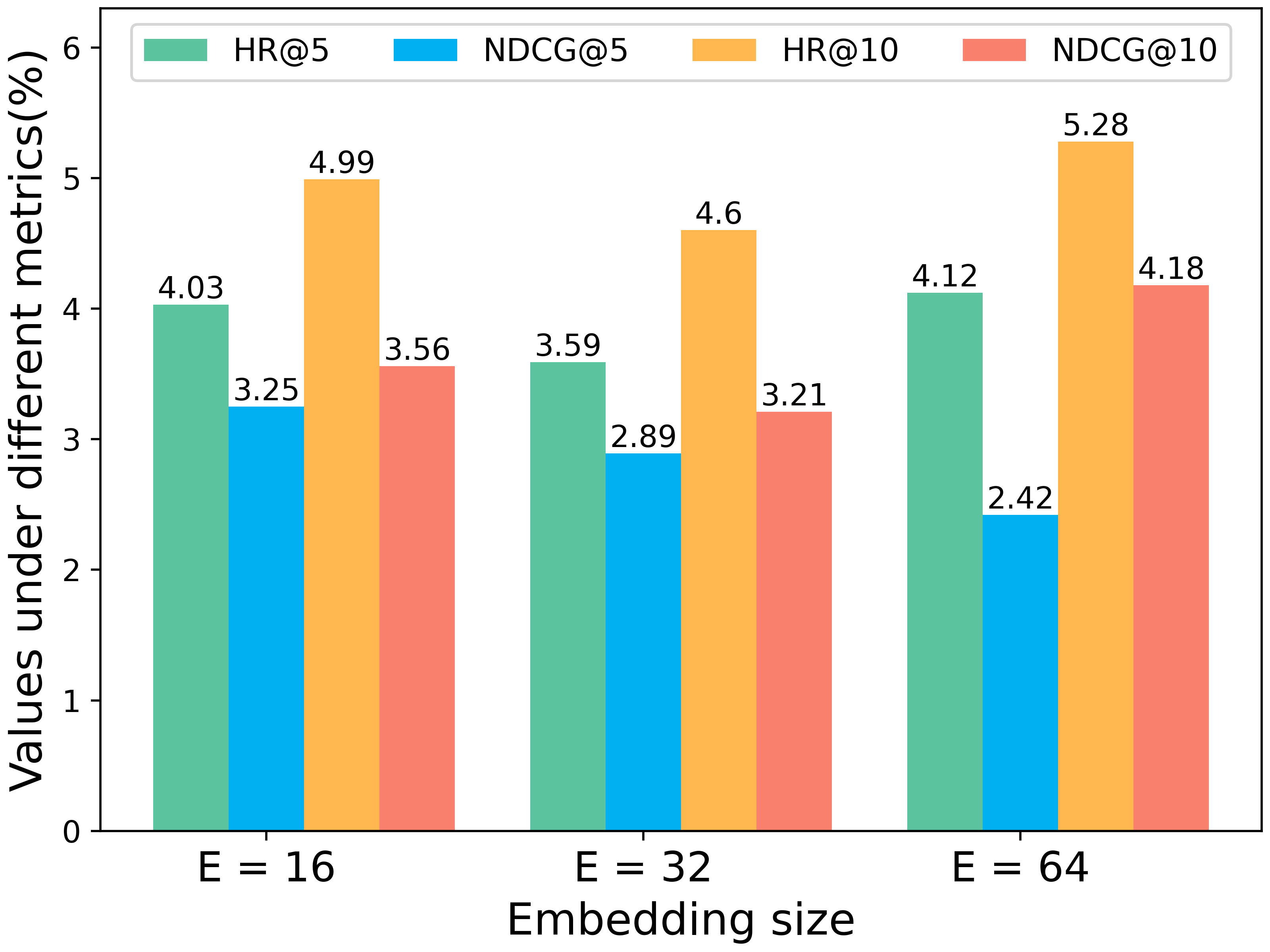}
        \caption{User-item coindexing}
        \label{fig:coui_gci_e}
    \end{subfigure}
    \hfill
\vspace{-15pt}
\end{figure*}

\begin{figure*}[htbp]
    \caption{Impact of the number of clusters $N$ used in KMeans clustering on the recommendation performance under the GCI settings, where \autoref{fig:u_gci_n}, \autoref{fig:i_gci_n}, \autoref{fig:ui_gci_n}, \autoref{fig:coui_gci_n} refers to user-only indexing, item-only indexing, useritem indexing, useritem-coindexing, respectively.}
    \label{fig:gci_n}
    \begin{subfigure}[h]{0.24\textwidth}
    \includegraphics[width=\textwidth]{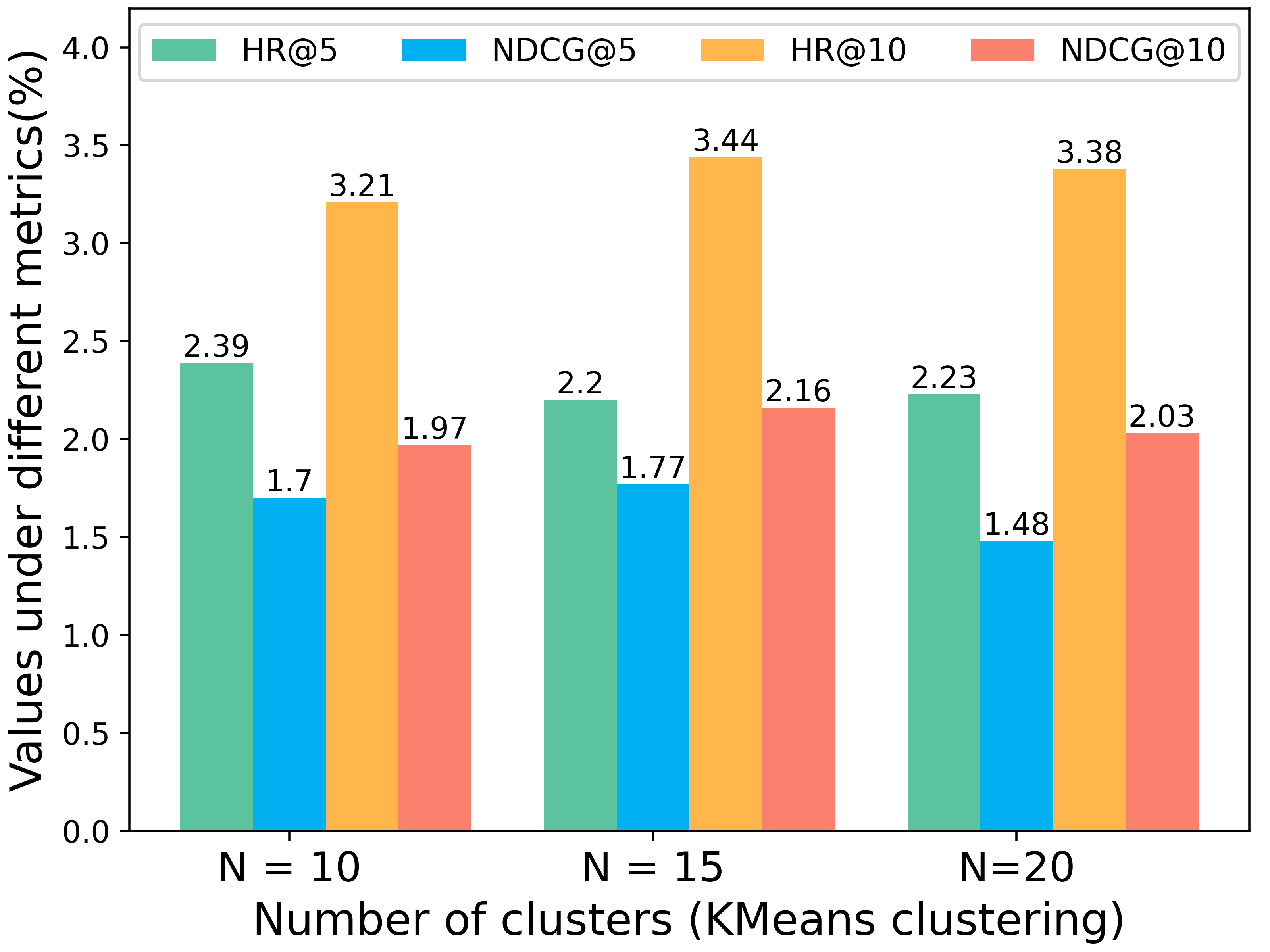}
        \caption{User-only indexing}
        \label{fig:u_gci_n}
    \end{subfigure}
    \hfill
    \begin{subfigure}[h]{0.24\textwidth}
    \includegraphics[width=\textwidth]{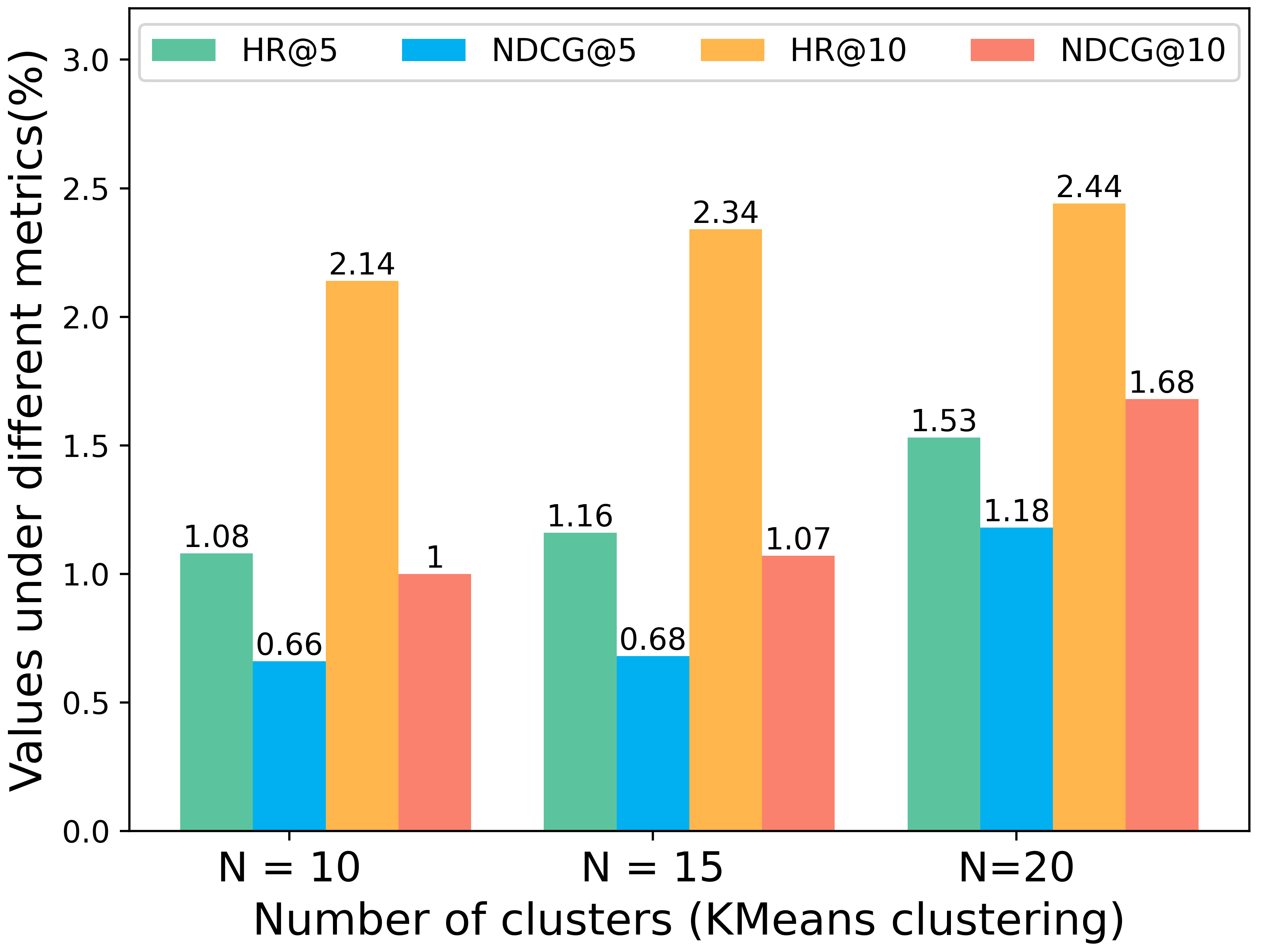}
        \caption{Item-only indexing}
        \label{fig:i_gci_n}
    \end{subfigure}
    \hfill
    \begin{subfigure}[h]{0.24\textwidth}
    \includegraphics[width=\textwidth]{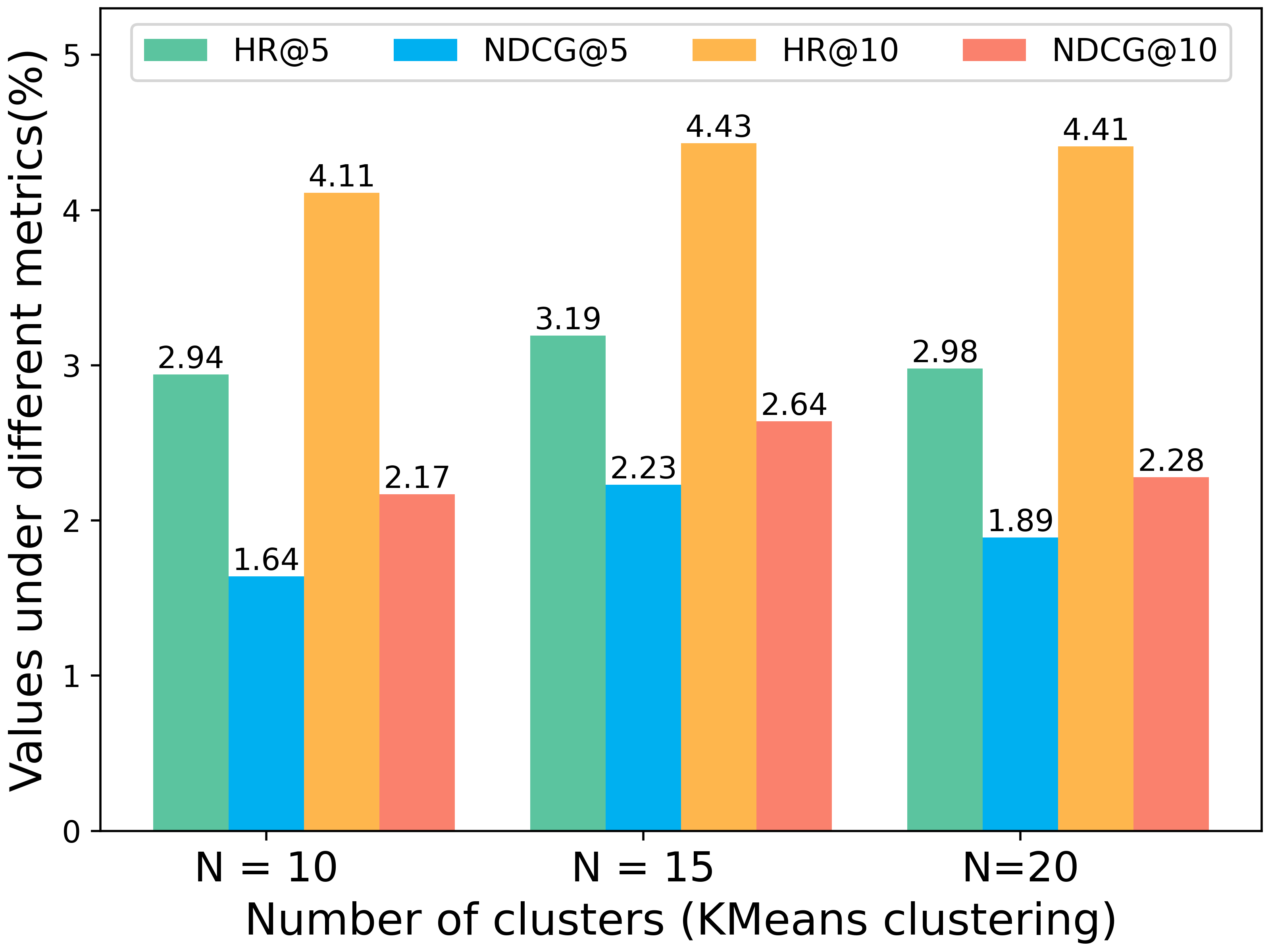}
        \caption{User-item indexing}
        \label{fig:ui_gci_n}
    \end{subfigure}
    \hfill
    \begin{subfigure}[h]{0.24\textwidth}
    \includegraphics[width=\textwidth]{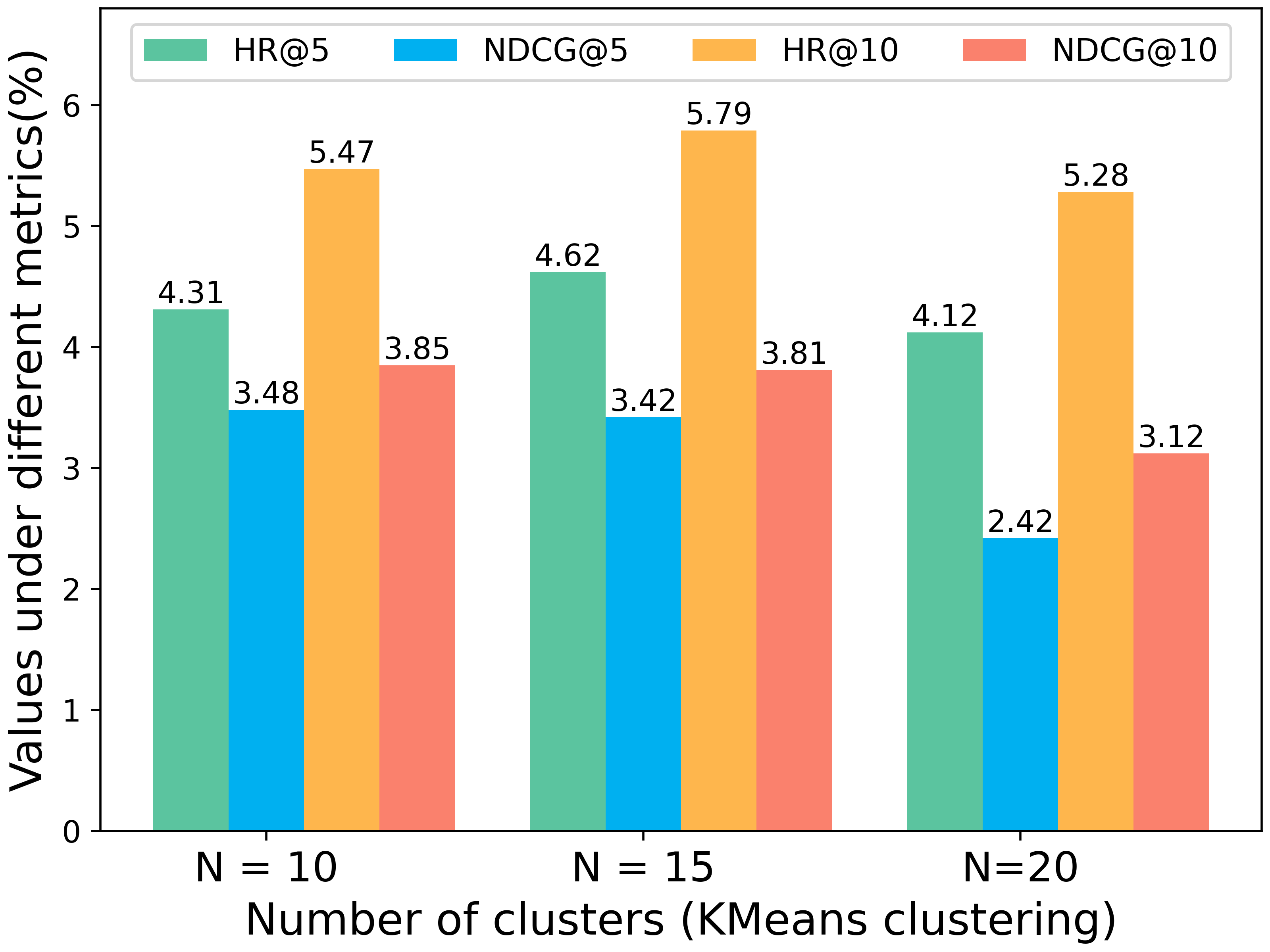}
        \caption{User-item coindexing}
        \label{fig:coui_gci_n}
    \end{subfigure}
    \hfill
\end{figure*}

In this part, we analyze the impact of two important parameters, i.e., embedding size $E$ and the number of clusters $N$ on the recommendation performance under the \textbf{GCI} settings. We use $E$ in [16, 32, 64] and evaluate the recommendation accuracy of \model/, respectively. The result is presented in \autoref{fig:gci_e}.
With the exception of GCI(CoUI), setting \(E\) to 32 and 64 generally yields superior recommendation performance for \model/ compared to when \(E\) is set to 16. This implies that graph embeddings require sufficient dimensional representation prior to quantization. However, the results obtained with \(E\) set to 32 are comparable to those with \(E\) set to 64, and even surpass them in the GCI(U) and GCI(UI) scenarios. This indicates that beyond a certain threshold for embedding size (e.g., 32), there may be no substantial enhancement in recommendation performance.
Similar to Section \ref{sec:impact_sci}, we also examine the effect of \(N\) on recommendation performance within the GCI configurations. However, for this study, we restrict \(N\) to the set \([10, 15, 20]\). This decision is informed by our observation that normalized embedding vectors within the GCI settings are more evenly clustered. Thus, if the value of \(N\) is too high, the collaborative ID tokens might become too truncated, losing valuable collaborative information.
From \autoref{fig:gci_n}, it's evident that when \(N\) is set to 15, \model/ achieves optimal results in GCI(U), GCI(UI), and GCI(CoUI). This aligns with our analysis in Section \ref{sec:impact_sci} which found that a more balanced number distribution benefits recommendation. 
Moreover, since the range of \(N\) values here is narrower than in the SCI scenario, the influence of \(N\) on recommendation accuracy is relatively subtle.

\vspace{-3pt}
\subsection{Efficiency Analysis}
In this section, we analyze the efficiency of \model/. For simplicity, we evaluate on the Toys dataset and only use \model/-SCI(CoUI) as a representative to compare. It is because \model/-SCI(CoUI) beats most of the baselines in terms of recommendation accuracy and the slight variation of other \model/ versions has subtle influences on the efficiency analysis.


\vspace{-3pt}
\subsubsection{Computational Efficiency}
We evaluate \model/ against the baselines discussed in Section \ref{sec:exp_setting}, considering both training epochs and total runtime (encompassing training and inference). As indicated in \autoref{tab:comp_efficiency}, \model/ stands out by necessitating the least number of training epochs (i.e., 8) compared to all other methods. Furthermore, the overall runtime of \model/ is second only to SimpleX. This underscores \model/'s computational efficiency and its ability to deliver satisfactory results without extensive training. Notably, \model/'s total runtime is just 0.9\% of P5's, marking a substantial enhancement. Thus, \model/ promises considerable time savings over other LM-based recommenders, while maintaining recommendation precision simutaneously.
\begin{table}[htbp]
\centering
\footnotesize
\setlength\tabcolsep{2pt}
\caption{Computation efficiency, where we compare \model/ with baselines on training epoches and overall runtime.} 
\label{tab:comp_efficiency}
\begin{tabular}{lccc}
\hline
Method                           & Training epoch                  &  & Overall runtime (mins)            \\ \hline
BPRMF                            & 122                             &  & 70.3                              \\
LightGCN                         & 99                              &  & 52.9                              \\
SimpleX                          & 21                              &  & \textbf{24.3}    \\
P5                               & \underline{20} &  & 1069.4                            \\
\model/ & \textbf{8}     &  & \underline{43.5} \\ \hline
\vspace{-15pt}
\end{tabular}
\end{table}

\vspace{-2pt}
\subsubsection{GPU Usage}
As \model/ differs from batch size and other parameters the baseline takes, we only compare the computation resource usage \model/ takes with P5, a state-of-the-art LM-based generative recommender. 
To ensure a fair comparison, we adopt identical settings, specifically, using the batch size of 64 during training and the batch size of 48 during inference, mirroring the settings taken with P5.  As shown in \autoref{tab:gpu_usage}, \model/ has fewer parameters than P5, with the smallest version containing only 59.06\% of the parameters found in P5. Moreover, \model/ can perform both training and inference on a single GPU, thereby utilizing significantly less GPU memory resources than P5. This efficiency is partly due to \model/'s use of only one prompt template, whereas P5 employs 11 prompt templates. Additionally, P5 requires sequential data to facilitate straightforward training, causing the input tokens to be much longer, as a single input will encompass multiple items. In contrast, \model/ focuses directly on straightforward tasks, thus consuming less GPU memory.

\section{Conclusion and Future Work} \label{sec:conclusion}
In this research, we present \model/, a lightweight LM-based recommender designed for straightforward generative recommendations. To enhance the model's ability of capturing collaborative signals between users and items, we introduce two advanced ID indexing techniques: Spectral Collaborative Indexing (SCI) and Graph Collaborative Indexing (GCI) for enhancing \model/'s recommendation performance.
\begin{table}[t]
\centering
\footnotesize
\setlength\tabcolsep{1pt}
\caption{GPU usage, where we compare \model/ under different inner dimension values ($w$), with P5.} \label{tab:gpu_usage}

\begin{tabular}{@{}lccccc@{}}
\toprule
Model                     & \#Param &  & GPU Usage(Training) &  & GPU Usage(Inference)\\ \midrule
P5                        & 60.75M    &  & 4GPUs (12.3GiB per GPU)        &  & 4GPUs (22.1GiB per GPU)             \\
\model/($w=16$) & 35.88M        &  & 1GPU (2.74GiB per GPU)     &  & 1GPU (7.35GiB per GPU)             \\
\model/($w=32$) & 36.06M        &  & 1GPU (2.75GiB per GPU)     &  & 1GPU (7.41GiB per GPU)             \\
\model/($w=64$) & 36.46M        &  & 1GPU (2.79GiB per GPU)     &  & 1GPU (7.47GiB per GPU)             \\ \bottomrule
\end{tabular}
\vspace{-12pt}
\end{table}
To address the issue of over-parameterization when using language models for recommendation, we tailor the Feed-Forward layers within the Transformer blocks, reducing the number of parameters without compromising performance.
Additionally, we optimize the conventional generation-then-retrieval pipeline by introducing a constrained generation approach to guarantee the existence of the generated items. This modification effectively tackles the hallucination problem in the generation process, ensuring more accurate recommendations.
Experiments conducted on three real-world datasets demonstrate that \model/ outperforms competitive baselines in terms of both recommendation accuracy and efficiency.
Concurrently, given that LightLM necessitates the construction of collaborative IDs, its efficacy in addressing the cold-start problem remains limited. Addressing this limitation will be a focal point of our subsequent research. Additionally, we have made an initial attempt at designing recommendation-tailored Transformer blocks in this work. Our future endeavors will involve investigating specialized Transformer architectures, such as recommendation-focused attention mechanisms. Such explorations aim to better cater to the distinctive demands of generative recommendation tasks, potentially resulting in more sophisticated and efficient recommender systems that can be applied across a wider range of applications.

\section*{Ethical Considerations}
Our method is proposed to enhance recommendation accuracy for users. Since our method does not involve privacy/safety problems, as long as it was applied appropriately, our approach can boost the efficiency of LM-based recommender systems without causing significant adverse societal effects.

\bibliographystyle{ACM-Reference-Format}
\bibliography{ref}



\end{document}